\documentclass{article}%
\usepackage{amsmath}
\usepackage{amsfonts}
\usepackage{amssymb}
\usepackage{graphicx}%
\newcommand{\bpartial}{\mathop{\partial\kern -4pt\raisebox{.8pt}{$|$}}}
\newcommand{\bra}{\mathopen{[\kern-1.6pt[}}
\newcommand{\ket}{\mathclose{]\kern-1.5pt]}}
\newcommand{\bbra}{\mathopen{[\kern-2.2pt[\kern-2.3pt[}}
\newcommand{\bket}{\mathclose{]\kern-2.1pt]\kern-2.3pt]}}

\makeindex
\begin{document}
\title{\bf Perturbed N=(2,2) supersymmetric sigma models on Lie groups }
\author {  M. Ebrahimi$^{a}$ \hspace{-2mm}{ \footnote{ e-mail: m$_{-}$ebrahimi@pnu.ac.ir}}\hspace{1mm},\hspace{1mm} A.
Rezaei-Aghdam$^{b}$
\hspace{-2mm}{ \footnote{Corresponding author. e-mail:
rezaei-a@azaruniv.edu}}\hspace{2mm}\\
{\small{\em $^{a}$Department of Physics, Faculty of science,
Payame Noor
University,}}\\ {\small{\em 19395-3697, Tehran, Iran  }}\\
{\small{\em
$^{b}$Department of Physics, Faculty of science, Azarbaijan Shahid Madani University }}\\
{\small{\em  53714-161, Tabriz, Iran  }}}

\maketitle
\begin{abstract}
We perturbed N=(2,2) supersymmetric WZW and sigma models on Lie
groups by adding a term to their actions. Then by using
non-coordinate basis we obtain conditions, from the algebraic
point of view, under which the N=(2,2) supersymmetry is preserved.
By applying this method, we have obtained conditions on the
existence of N=(2,2) supersymmetry on the Drinfeld action (master
action for the Poisson-Lie T-dual sigma models).

\end{abstract}
\newpage
\section{\bf Introduction}
String theories with N=2 worldsheet supersymmetry give rise to
spacetime physics which is itself supersymmetric. Spacetime
supersymmetry is our best hope for addressing the hierarchy
problem and hence has strong theoretical motivation. Furthermore,
supersymmetric sigma models are of interest,e.g, as gauge-fixed
actions, as representing exact string vacuum (WZW model), for
their intimate connection to complex geometry of target manifold
\cite{Gat}, \cite{sev} and for their role as effective low-energy
actions for supergravity scalars. From the geometrical point of
view the N=(2,2) extended supersymmetry in sigma model is
equivalent to the existence of bi-Hermitian structure on the
target manifold such that the complex structures are covariantly
constant with respect to torsionful affine connections \cite{Gat}
(see also \cite{Lya} and references therein ). Furthermore, it is
shown that the algebraic structures related to these bi-Hermitian
relations for the N=(2,2) supersymmetric WZW models are the Manin
triples \cite{Par}, \cite{Lin}. Meanwhile the algebraic structure
associated to the bi-Hermitian geometry for N=(2,2) supersymmetric
sigma models on Lie groups are found recently in \cite{RS}. On the
other hand, T-duality is the most important symmetries of string
theory \cite{Bu}. In this way, Poisson-Lie T-duality, a
generalization of T-duality, does not require existence of
isometry in the original target manifold (as in usual
T-duality\cite{Bu}), \cite{Kl}, \cite{Ks}. There are some attempts to find
the effect of T-duality on the N=(2,2) supersymmetry. In \cite{Sh}
and \cite{Fh}, it is shown that the N=(2,2) supersymmetry is
preserved under Abelian T-duality; both in the cases where complex
structures are independent and dependent on the coordinates to
which the T-duality is performed . In the dependency case it is
shown that  the extended worldsheet supersymmetry is non-local
under T-duality transformation \cite{Fh}. In \cite{Hj} there are some attempts for
studing the effect of Poisson-Lie T-duality on N=(2,2)
supersymmetry. Here, we try to have one step  forward in this
direction by studying conditions on the existence of N=(2,2)
supersymmetry on the Drinfeld action(master action for the
 Poisson-Lie T-dual sigma models). The paper is organized as
follows. \par In section two, to introduce the notations and
selfcontaing the paper we review the N= (2,2) supersymmetric WZW
and sigma models on Lie groups from geometrical and algebraic
point of view. Then, in section three we first perturb N=(2,2)
supersymmetric WZW and sigma models on Lie groups by adding a general term
to their action; then by using of non-coordinate bases we obtain
from the algebraic point of view, conditions under which the
N=(2,2) supersymmetry is preserved. In section four we obtain
conditions on the existence of N=(2,2) supersymmetry on Drinfeld
action (master action for the Poisson-Lie T-dual sigma
models)\cite{Ks}.

\section{\bf Review of $N=(2,2)$ supersymmetric WZW and sigma models } 
In this section we review the results of N=(2,2)
supersymmetric sigma models on the manifolds \cite{Gat} in general and on the Lie groups as special case (see for example \cite{Lya} ,\cite{Lin}). We start from the
general N=1 supersymmetric sigma models action on the manifold M
as follows:

\begin{equation}
S=\int d^{2}\sigma d^{2}\theta
D_{+}\Phi^{\mu}D_{-}\Phi^{\nu}(G_{\mu\nu}(\Phi)+B_{\mu\nu}(\Phi)),
\end{equation}
where $\Phi^{\mu}$are N=1 superfields with bosonic parts as
coordinates of the manifold M, meanwhile the bosonic parts of
$G_{\mu\nu}$ and $B_{\mu\nu}$ are respectivly metric and
antisymmetric tensor on M. The action (1) is manifestly invariant
under supersymmetry transformations
\begin{equation}
\delta^{1}(\epsilon)\Phi^{\mu}=i(\epsilon^{+}Q_{+}+\epsilon^{-}Q_{-})\Phi^{\mu},
\end{equation}
furthermore this action has additional non manifest supersymmetry
of the form

\begin{equation}
\delta^{2}(\epsilon)\Phi^{\mu}=\epsilon^{+}D_{+}\Phi^{\nu}J^{\mu}_{+\nu}(\Phi)+\epsilon^{-}D_{-}\Phi^{\nu}J^{\mu}_{-\nu}(\Phi),
\end{equation}
where in the above relations $Q_{\pm}$ and $D_{\pm}$ are
supersymmetry generators and superderivatives respectively furthermore $\epsilon^{\pm}$ are parameters of supersymmetry transformations and
$J^{\rho}_{\pm\sigma} \in TM \bigotimes T^{\ast}M$. Invariance of
the action (1) under the transformations (3) imposes
 the following conditions on $J^{\rho}_{\pm\sigma}$:
\begin{equation}
J^{\mu}_{\pm\lambda}J^{\lambda}_{\pm\nu}=-\delta^{\mu}_{\nu},
\end{equation}

\begin{equation}
\hspace{1cm}J^{\mu}_{\pm\rho}G_{\mu\nu}=-G_{\mu\rho}J^{\mu}_{\pm\nu},
\end{equation}

\begin{equation}
\nabla^{(\pm)}_{\rho}J^{\mu}_{\pm\nu}=J^{\mu}_{\pm\nu,\rho}+\Gamma^{\pm\mu}_{\rho\sigma}J^{\sigma}_{\pm\nu}-\Gamma^{\pm\sigma}_{\rho\nu}J^{\mu}_{\pm\sigma}=0,
\end{equation}
where

\begin{equation}
\Gamma^{\pm\mu}_{\rho\nu}=\Gamma^{\mu}_{\rho\nu}\pm
G^{\mu\sigma}H_{\sigma\rho\nu},
\end{equation}
such that $\Gamma^{\mu}_{\rho\nu}$ is the usual Christoffel symbols
and H being the torsion three form

\begin{equation}
H_{\mu\rho\sigma}=\frac{1}{2}(B_{\mu\rho,\sigma}+B_{\rho\sigma,\mu}+B_{\sigma\mu,\rho}),
\end{equation}

 In order to have a closed on-shell
supersymmetry algebra with generators (2) and (3) we must have
zero Nijenhuis tensor \cite{Ni} for $J^{\mu}_{\pm\nu}$ \cite{Gat}, \cite{sev} ; 

\begin{equation}
N^{\rho}_{\mu\nu}(J_{\pm})=J^{\gamma}_{\pm\mu}\partial_{[\gamma}J^{\rho}_{\pm\nu]}-J^{\gamma}_{\pm\nu}\partial_{[\gamma}J^{\rho}_{\pm\mu]}=0.
\end{equation}
In this way, having an N=(2,2) supersymmetric sigma models on the
manifold M is geometrically equivalent to have two bi-Hermitian
complex structure $ J_{\pm}$ such that their covariant derivations
with respect to connections $\Gamma^{\pm\mu}_{\rho\nu}$ are equal
to zero (6). The vanishing of the Nijenhuis tensor (9) (the
integrability condition) and condition (6) implys that the complex
structures $ J_{\pm}$ should preserve the torsion [1-3]; i.e

\begin{equation}
H_{\delta\nu\lambda}=J^{\sigma}_{\pm\delta}J^{\rho}_{\pm\nu}H_{\sigma\rho\lambda}+J^{\sigma}_{\pm\lambda}J^{\rho}_{\pm\delta}H_{\sigma\rho\nu}+J^{\sigma}_{\pm\nu}J^{\rho}_{\pm\lambda}H_{\sigma\rho\delta}.
\end{equation}
In the case that M is a Lie group G, then using non-coordinate
bases, we have
\begin{equation}
G_{\mu\nu}=L_{\mu}^{A}L_{\nu}^{B}G_{AB}=R_{\mu}^{A}R_{\nu}^{B}G_{AB},
\end{equation}
\begin{equation}
f_{AB}\hspace{0cm}^{C}=L^{C}\hspace{0cm}_{\nu}(L_{A}\hspace{0cm}^{\mu}\partial_{\mu}L_{B}\hspace{0cm}^{\nu}-L_{B}\hspace{0cm}^{\mu}\partial_{\mu}L_{A}\hspace{0cm}^{\nu})=
R^{C}\hspace{0cm}_{\nu}(R_{A}\hspace{0cm}^{\mu}\partial_{\mu}R_{B}\hspace{0cm}^{\nu}-R_{B}\hspace{0cm}^{\mu}\partial_{\mu}R_{A}\hspace{0cm}^{\nu}),
\end{equation}

\begin{equation}
H_{\mu\nu\lambda}=L_{\mu}^{A}L_{\nu}^{B}L_{\lambda}^{C}H_{ABC}=R_{\mu}^{A}R_{\nu}^{B}R_{\lambda}^{C}H_{ABC},
\end{equation}

\begin{equation}
J^{\mu}_{-\nu}=L^{\mu}_{A}J^{A}\hspace{0cm}_{B}L_{\nu}^{B},\hspace{2cm}J^{\mu}_{+\nu}=R^{\mu}_{A}J^{A}\hspace{0cm}_{B}R_{\nu}^{B},
\end{equation}
where $G_{AB}$ is symmetric ad-invariant non-degenerat bilinear
form and $H_{ABC}$ is antisymmetric tensor on Lie algebra $\bf g$
\footnote{ Note that $G_{\mu\nu}(G_{AB})$raise and lower the
target space (Lie algebra) indices.}; furthermore
$L_{\mu}^{A}(R_{\mu}^{A})$ and $L^{\mu}_{A}(R^{\mu}_{A})$ are left
(right) invariant veilbien and their inverses on the Lie group G
respectively,  and $f_{AB}\hspace{0cm}^{C}$is the structure
constant of Lie algebra and J is a Lie algebraic map; $ J: \bf g
\longrightarrow \bf g $. In this case  by use of (11-14) the relations
(4-6), (9) and (10)
 can be rewritten in the following algebraic form
\cite{RS}:
\begin{equation}
J^{2}=-I,\hspace{1mm}
\end{equation}

\begin{equation}
{\chi}_{A}+J^{t}\hspace{1mm}{\chi}_{A}\hspace{1mm}J^{t}+J^{B}\hspace{0cm}_{A}\hspace{1mm}{\chi}_{B}\hspace{1mm}J^{t}-J^{B}\hspace{0cm}_{A}\hspace{1mm}J^{t}\hspace{1mm}{\chi}_{B}=0,
\end{equation}

\begin{equation}
J^{t}\hspace{1mm}G\hspace{1mm}J=G,
\end{equation}

\begin{equation}
H_{A}= J^{t} (H_{B} {J^{B}}\hspace{0cm}_{A}) + J^{t}H_{A}J+(H_{B}
{J^{B}}\hspace{0cm}_{A}) J,
\end{equation}

\begin{equation}
J^{t}(H_{A}+\chi_{A}G) =(J^{t}(H_{A}+\chi_{A}G))^{t}.
\end{equation}
where  ${(\chi_A)_B}^C=-{f_{AB}}^C$ is the adjoint representation
and $(H_A)_{BC}=H_{ABC}$ ; $G_{AB}$is the ad-invariant metric on
Lie algebra $\bf g$, such that:
\begin{equation}
(\chi_A  G)^{t}=-\chi_A  G.
\end{equation}
 These relations shows that N=(2,2)
supersymmetric sigma models on Lie groups $G$ have a geometric
biHermitian structures on the Lie group $G$ \cite{Gat} or
equivalently an algabraic bi-Hermitian structures $(J,G, H)$ on
Lie algebra $\bf g$ \cite{RS}.

For N=2 supersymmetric WZW models on the Lie group G we have (up
to constant)

\begin{equation}
H_{\mu\nu\lambda}=L_{\mu}\hspace{0cm}^{A}L_{\nu}\hspace{0cm}^{B}L_{\lambda}\hspace{0cm}^{C}f_{ABC}=R_{\mu}\hspace{0cm}^{A}R_{\nu}\hspace{0cm}\hspace{0cm}^{B}R_{\lambda}^{C}f_{ABC},
\end{equation}
i.e. $H_{ABC}= f_{ABC}$. In this case relation (16) shows that we
have a Lie bialgebra structures on $\bf g$ \cite{Par}; and
relation (18) reduce to (16) and (19) is automatically satisfied
i.e. Lie bialgebra structure is a special case of algebraic
biHermitian structure $(J,G,H)$ with $H_{ABC}=f_{ABC}$ \cite{RS}.

\section{\bf Perturbed N=(2,2) supersymmetric WZW and sigma models on Lie groups.}
In this section we find conditions such that the perturbed N=(2,2)
supersymmetric WZW and sigma models on Lie groups preserve N=(2,2)
supersymmetry. We assume that the action (1) (as sigma models on
Lie group or WZW model) has N=(2,2) supersymmetry, and is
perturbed with the following general term:

\begin{equation}
S'=\int d^{2}\sigma d^{2}\theta
D_{+}\Phi^{\mu}D_{-}\Phi^{\nu}(G'_{\mu\nu}(\Phi)+B'_{\mu\nu}(\Phi)),
\end{equation}
where the bosonic part of $G'_{\mu\nu}$ and $B'_{\mu\nu}$ are
symmetric and antisymmetric tensors on the Lie group G. In this
case we have the following action:

\begin{equation}
S^{''}=S+S'=\int d^{2}\sigma d^{2}\theta
D_{+}\Phi^{\mu}D_{\nu}\Phi^{\nu}(G^{''}_{\mu\nu}(\Phi)+B^{''}_{\mu\nu}(\Phi)),
\end{equation}
 such that $G^{''}_{\mu\nu}=G_{\mu\nu}+G'_{\mu\nu}$and
$B^{''}_{\mu\nu}=B_{\mu\nu}+B'_{\mu\nu}$or
$H^{''}_{\mu\nu\lambda}=H'_{\mu\nu\lambda}+H_{\mu\nu\lambda}$. Now
we find conditions under which the action $S^{''}$have N=(2,2)
supersymmetry; i.e. the relations (4)-(6) and (10) holds for
$G^{''}_{\mu\nu}$ and $H^{''}_{\mu\nu\lambda}$\footnote{Note that
relation (9) is equivalent to relation (10)}. Similar to the
previous section we use (11)-(14) and the following relations:

\begin{equation}
G'_{\mu\nu}=L_{\mu}\hspace{0cm}^{A}L_{\nu}\hspace{0cm}^{B}G'_{AB},\hspace{1cm}
G^{''}_{\mu\nu}=L_{\mu}\hspace{0cm}^{A}L_{\nu}\hspace{0cm}^{B}G^{''}_{AB},
\end{equation}
\vspace{.1cm}
\begin{equation}
H'_{\mu\nu\lambda}=L_{\mu}\hspace{0cm}^{A}L_{\nu}\hspace{0cm}^{B}L_{\lambda}\hspace{0cm}^{C}H'_{ABC}=R_{\mu}\hspace{0cm}^{A}R_{\nu}\hspace{0cm}\hspace{0cm}^{B}R_{\lambda}\hspace{0cm}^{C}H'_{ABC},
\end{equation}

\begin{equation}
H^{''}_{\mu\nu\lambda}=L_{\mu}\hspace{0cm}^{A}L_{\nu}\hspace{0cm}^{B}L_{\lambda}\hspace{0cm}^{C}H^{''}_{ABC}=R_{\mu}\hspace{0cm}^{A}R_{\nu}\hspace{0cm}\hspace{0cm}^{B}R_{\lambda}\hspace{0cm}^{C}H^{''}_{ABC},
\end{equation}
with the assumptions that $G_{AB}$, $G'_{AB}$ and $G^{''}_{AB}$
are constant (independent from coordinates of Lie group G and its
Lie algebra $\bf g$) and invertible. Now by use of the above
relations one can discussed about perturbed algebraic bi-Hermitian
structure instead of perturbed $N=(2,2)$ supersymmetry. We
perform these in the following seven cases\footnote{Note that the five cases b)-g) are especial case of a), but we discusses them for further details}:\\

{\bf Case a)} The algebraic bi-Hermitian structure $(J,G,H)$ is
perturbed with $(J',G',H')$. Now we must impose the following
conditions on $(J',G',H')$ such that the structure$(J'',G'',H'')$
become
bi-Hermitian (i.e.relations (15)-(20) holds for $(J'',G'',H'')$).\\
From relation (15) we obtain:
\begin{equation}
\{J',J+J'/2\}=0,
\end{equation}
where in the case that $J'$ is a complex structure we have
\begin{equation}
\{J,J'\}=1 .
\end{equation}
From relation (16) we obtain
\begin{equation}
J^{t}\hspace{1mm}{\chi}_{A}\hspace{1mm}J'^{t}+J'^{t}\hspace{1mm}{\chi}_{A}\hspace{1mm}J^{t}+J'^{t}\hspace{1mm}{\chi}_{A}\hspace{1mm}J'^{t}
+J^{B}\hspace{0mm}_{A}\hspace{1mm}{\chi}_{B}\hspace{1mm}J'^{t}+J'^{B}\hspace{0mm}_{A}\hspace{1mm}{\chi}_{B}\hspace{1mm}J^{t}+
 J'^{B}\hspace{0mm}_{A}\hspace{1mm}{\chi}_{B}\hspace{1mm}J'^{t}
\end{equation}
$$
 -J^{B}\hspace{0mm}_{A}\hspace{1mm}J'^{t}\hspace{1mm}{\chi}_{B}-
 J'^{B}\hspace{0mm}_{A}\hspace{1mm}J^{t}\hspace{1mm}{\chi}_{B}-J'^{B}\hspace{0mm}_{A}\hspace{1mm}J'^{t}\hspace{1mm}{\chi}_{B}=0,
$$\\
in the case that $J'$ is a complex structure we have
\begin{equation}
N(J,J')=J^{t}\hspace{1mm}{\chi}_{A}\hspace{1mm}J'^{t}+J'^{t}\hspace{1mm}{\chi}_{A}\hspace{1mm}J^{t}+
+J^{B}\hspace{0mm}_{A}\hspace{1mm}{\chi}_{B}\hspace{1mm}J'^{t}+J'^{B}\hspace{0mm}_{A}\hspace{1mm}{\chi}_{B}\hspace{1mm}J^{t}
-J^{B}\hspace{0mm}_{A}\hspace{1mm}J'^{t}\hspace{1mm}{\chi}_{B}-
 J'^{B}\hspace{0mm}_{A}\hspace{1mm}J^{t}\hspace{1mm}{\chi}_{B}-{\chi}_{A}=0,
\end{equation}\\
where $N(J,J')$ is the \emph{Nijenhuis concomitant} of J and $J'$
\cite{Fn}, \cite{Ni}.
 Furthermore, one can obtain the following
relation from (17):
\begin{equation}
J^{t}\hspace{1mm}G\hspace{1mm}J'+J^{t}\hspace{1mm}G'\hspace{1mm}J+J^{t}\hspace{1mm}G'\hspace{1mm}J'+J'^{t}\hspace{1mm}G\hspace{1mm}J+
J'^{t}\hspace{1mm}G\hspace{1mm}J'+J'^{t}\hspace{1mm}G'\hspace{1mm}J+J'^{t}\hspace{1mm}G'\hspace{1mm}J'=G',
\end{equation}
where in the case that $(J',G')$ is bi-Hermitian  complex
structure we have
\begin{equation}
J^{t}\hspace{1mm}G\hspace{1mm}J'+J^{t}\hspace{1mm}G'\hspace{1mm}J+J^{t}\hspace{1mm}G'\hspace{1mm}J'+J'^{t}\hspace{1mm}G\hspace{1mm}J+
J'^{t}\hspace{1mm}G\hspace{1mm}J'+J'^{t}\hspace{1mm}G'\hspace{1mm}J=0.
\end{equation}
Finally from (18) and (19) in the case that $(J',G',H')$ is
algebraic bi-Hermitian structure, we obtain the following
relations respectively:
\begin{equation}
J^{t}\hspace{1mm}H_{B}\hspace{1mm}J'^{B}\hspace{0mm}_{A}
+J^{t}\hspace{1mm}H'_{B}\hspace{1mm}J^{B}\hspace{0mm}_{A}+J^{t}\hspace{1mm}H'_{B}\hspace{1mm}J'^{B}\hspace{0mm}_{A}
+J'^{t}\hspace{1mm}H_{B}\hspace{1mm}J^{B}\hspace{0mm}_{A}+J'^{t}\hspace{1mm}H_{B}\hspace{1mm}J'^{B}\hspace{0mm}_{A}+J'^{t}\hspace{1mm}H'_{B}\hspace{1mm}J^{B}\hspace{0mm}_{A}
\end{equation}
$$
+J^{t}\hspace{1mm}H_{A}\hspace{1mm}J'
+J^{t}\hspace{1mm}H'_{A}\hspace{1mm}J+J^{t}\hspace{1mm}H'_{A}\hspace{1mm}J'+J'^{t}\hspace{1mm}H_{A}\hspace{1mm}J+J'^{t}\hspace{1mm}H_{A}\hspace{1mm}J'
+J'^{t}\hspace{1mm}H'_{A}\hspace{1mm}J+H_{B}\hspace{1mm}J^{B}\hspace{0mm}_{A}J'
$$
$$
+H_{B}\hspace{1mm}J'^{B}\hspace{0mm}_{A}J+H_{B}\hspace{1mm}J'^{B}\hspace{0mm}_{A}J'
+H'_{B}\hspace{1mm}J^{B}\hspace{0mm}_{A}J+H'_{B}\hspace{1mm}J^{B}\hspace{0mm}_{A}J'
+H'_{B}\hspace{1mm}J'^{B}\hspace{0mm}_{A}J=0,
$$

\begin{equation}
J^{t}(H'_{A}+\chi_{A}G')+J'^{t}(H_{A}+\chi_{A}G)=-(H'_{A}+\chi_{A}G')J-(H_{A}+\chi_{A}G)J'.
\end{equation}
Meanwhile from (20) we see that $G'$ must be ad-invariant metric i.e.\\
\begin{equation}
\chi_{A}G'=-(\chi_{A}G')^{t}.
\end{equation}

{\bf Case b)} The algebraic bi-Hermitian structure $(J,G,H)$ is
perturbed with $(0,G',H')$. We see that this case is special case
of the above case. So from above relations we see that if
$(J,G',H')$ is an algebraic bi-Hermitian structure, then
$(J,G'',H'')$ is an
algebraic bi-Hermitian structure.\\

{\bf Case c)} The algebraic bi-Hermitian structure $(J,G,H)$ is
perturbed with $(J',0,H')$.This case is a special case of case a)
when $G'=0$. So relations (27),(29) are also satisfied for this
case and instead of (31) we must have
\begin{equation}
J^{t}\hspace{1mm}G\hspace{1mm}J'+J'^{t}\hspace{1mm}G\hspace{1mm}J+
J'^{t}\hspace{1mm}G\hspace{1mm}J'=0.
\end{equation}
Furthermore, relation (33) is also must be imposed on $(J',G,H')$
and instead of (34) we must have
\begin{equation}
J^{t}H'_{A}+J'^{t}(H_{A}+\chi_{A}G)+J'^{t}H'_{A}=-H'_{A}J-(H_{A}+\chi_{A}G)J'-H'_{A}J',
\end{equation}
such that by assuming $(J',G,H')$ is a bi-Hermitian structure, instead of (36) and (37) we have
$$
J^{t}\hspace{1mm}G\hspace{1mm}J'+(J^{t}\hspace{1mm}G\hspace{1mm}J')^t=-G,
$$
$$
(H'_{A}J+H_{A}J')^t=H'_{A}J+H_{A}J'.
$$

{\bf Case d)} The algebraic bi-Hermitian structure $(J,G,H)$ is
perturbed with $(J',G',0)$. This case is a special case of case a)
when $H'=0$. By assuming $J'^{2}=-1$ and that $(J',G,H')$ is a bi-Hermitian structure, the relations (27)-(32) are
also satisfied for this case and instead of (33) and (34) we have:
\begin{equation}
J^{t}H_{B}J'^{B}\hspace{0mm}_{A}+J'^{t}H_{B}J^{B}\hspace{0mm}_{A}+J^{t}H_{A}J'+J'^{t}H_{A}J+H_{B}J^{B}\hspace{0mm}_{A}J'
-H_{B}J'^{B}\hspace{0mm}_{A}J+H_{A}=0,
\end{equation}
\begin{equation}
J'^{t}\chi_{A}G-(J'^{t}\chi_{A}G)^{t}=-(J^{t}\chi_{A}G'-(J^{t}\chi_{A}G')^t),
\end{equation}
and relation (35) is also must be satisfied for this case.\\

{\bf Case e)} The algebraic bi-Hermitian structure $(J,G,H)$ is
perturbed with $(J',0,0)$. For this case relations (27)-(30) must
be satisfied and by assuming that $(J',G,H)$ is a bi-Hermitian structure then relation (34) is automatically satisfied and instead of (33) relation (38) must be satisfied and instead of (31)we must have
\begin{equation}
J^{t}GJ'+(J^{t}GJ')^{t}=-G.
\end{equation}
i.e. for having $(J'',G,H)$ as a bi-Hermitian structure we must
have $(J',G,H)$ as bi-Hermitian structure such that relations (28),(30),(38)and (40) must be satisfied. \\

{\bf Case f)} The algebraic bi-Hermitian structure $(J,G,H)$ is
perturbed with $(0,G',0)$. In this case from (17) we have
\begin{equation}
J^{t}G'J=G',
\end{equation}
and from (19) we obtain:
\begin{equation}
(\chi_{A}G'J)^{t}=\chi_{A}G'J,
\end{equation}
furthermore from (20) we have
\begin{equation}
\chi_{A}G'=-(\chi_{A}G')^{t},
\end{equation}
i.e. for obtaining $(J,G'',H)$ as an algebraic bi-Hermitian
structure we must have $G'$ as an ad-invariant metric such that
$(J,G',H)$ is also bi-Hermitian complex structure and the matrix
$\chi_{A}G'J$
must be symmetric matrix.\\

{\bf Case g)} The algebraic bi-Hermitian structure $(J,G,H)$ is
perturbed with $(0,0,H')$. From (19) we obtain
\begin{equation}
(H'_{A}J)^{t} =H'_{A}J.
\end{equation}
Note that in this case by assuming that $(J,G,H')$ is bi-Hermitian structure then from relation (18) we can not obtain a new
result. In this way, for having $(J,G,H'')$ as an algebraic
bi-Hermitian structure we must have $H'_{A}J$ as a symmetric
matrix.\\

\section{\bf Conditions on the existence of N=(2,2) supersymmetry
on Drinfeld action } In this section, we consider Drinfeld super
action as an example of perturbed N=(2,2) supersymmetric WZW
model. The form of Drinfeld super action as an action on Drinfeld
Lie group D is as follow \cite{SF}:
\begin{equation}
S=I_{0}(L)+ S',
\end{equation}
with
\hspace{5cm}$S'=-\frac{1}{2\pi}\int \langle
L^{-1}D_{+}L|R|L^{-1}D_{-}L\rangle
d^{2}\sigma d^{2}\theta$,\\
\\
where $I_{0}(L)$ is the N=1 supersymmetric WZW action and L is an
extension of Lie group element such that its bosonic part is $l\in
D$\cite{SF}; furthermore the operator R in terms of bases $R^{\pm}$ has the
following form \cite{SF}:
\begin{equation}
R=|R^{+}_{a}\rangle \eta^{ab} \langle R^{+}_{b}|+|R^{-}_{a}\rangle
\eta^{ab} \langle R^{-}_{b}|,
\end{equation}
such that :
\begin{equation}
\langle R^{\pm}_{a} | R^{\pm}_{b}\rangle, \langle R^{+}_{a} |
R^{-}_{b}\rangle=0,
\end{equation}
\begin{equation}
|R^{+}_{a}\rangle \eta^{ab} \langle R^{+}_{b}|-|R^{-}_{a}\rangle
\eta^{ab} \langle R^{-}_{b}|=I,
\end{equation}
\begin{equation}
R^{\pm}_{a}=T_{a} \pm (E^{\pm}_{0})_{ab} \tilde{T}^{b}
\hspace{1mm}, \eta_{ab}=(E^{+}_{0})_{ab}+(E^{-}_{0})_{ab},
\end{equation}
where $\{T_{a}\}$ and $\{\tilde{T}^{a}\}$ are the bases of the Lie
algebra $\bf g$ and $\bf{\tilde{g}}$ such that $\cal D=\bf g
\bigoplus \bf{\tilde{g}}$  is a Lie algebra of Drinfeld double
$\bf D$. The matrix $E_{0}^{+}$ is a arbitrary constant matrix and
$E_{0}^{-}$ is its transpose. Note that the action (45) is the
master action for the Poisson-Lie T-dual sigma models
\cite{Ks}. Indeed for the decompositions
$L=g\tilde{h}$ ($L=\tilde{g}h$)one can obtain the sigma model (and
its T-dual) as follows:
\begin{equation}
S=\int[(E_{0}^{\pm})^{-1}\pm\Pi(g)]_{ij}^{-1}(\partial_{+}gg^{-1})^{i}(\partial_{-}gg^{-1})^{j}
d\xi^{+} d\xi^{-}d^{2}\theta
\end{equation}
and
\begin{equation}
\tilde{S}=\int[E_{0}^{\pm}\pm\tilde{\Pi}(\tilde{g})]_{ij}^{-1}(\partial_{+}\tilde{g}\tilde{g}^{-1})^{i}(\partial_{-}\tilde{g}\tilde{g}^{-1})^{j}
d\xi^{+} d\xi^{-}d^{2}\theta
\end{equation}
with
\begin{equation}
\Pi(g)=b(g)a^{-1}(g)
\end{equation}
such that:
\begin{equation}
g^{-1}T_{a}g=a(g)_{a}\hspace{0cm}^{b}T_{b},\hspace{3mm}g^{-1}\tilde{T}^{a}g=b(g)^{ab}T_{b}+(a^{-1}(g))_{b}\hspace{0cm}^{a}\tilde{T}^{b}
\end{equation}
and in the same way for $\tilde{\Pi}(\tilde{g})$,  $\tilde{g}^{-1}T_{a}\tilde{g}$ and $\tilde{g}^{-1}\tilde{T}^{a}\tilde{g}$.\\
 Note that in this case the WZW action $I_{0}(L)$ has $N=(2,2)$
superysymmetry, because we have Lie bialgebra structure $(\bf
g,\bf{\tilde{g}})$\cite{Lin}. Now using above relation and using
$L^{-1}D_{\pm}L=L_{\mu}\hspace{0cm}^{A}D_{\pm}X^{\mu}T_A$ where $
X^{\mu}$ are superfield with bosonic sections as coordinates of
Lie group $\bf D$; and by use of the following decomposition for
$\cal D$:
\begin{equation}
L_{\mu}\hspace{0cm}^{A}T_{A}=L_{\mu}\hspace{0cm}^{a}T_{a}+L_{\mu,n+a}\tilde{T}^{a}
\end{equation}
and using of isotropy condition on inner product i.e.
\begin{equation}
<T_{a},T_{b}>=<\tilde{T}^{a},\tilde{T}^{b}>=0, \hspace{2mm}<T_{a},\tilde{T}^{b}>=\delta_{a}\hspace{0cm}^{b}
\end{equation}
the perturbed terms can be rewritten as follows:
\begin{equation}
S'=-\frac{1}{2\Pi}\int
L_{\mu}\hspace{0cm}^{A}E'_{AB}L_{\nu}\hspace{0cm}^{B}
D_{+}X^{\mu}D_{-}X^{\nu}d^{2}\sigma d^{2}\theta,
\end{equation}
such that the background matrix $E'_{AB}$ has the following form :
\begin{equation}
E^{'}_{AB}=\left( \begin{tabular}{c c}
                 $E^{-}_{0}\eta^{-1}E^{+}_{0}+E^{+}_{0}\eta^{-1}E^{-}_{0}$ &
                 $-(E^{+}_{0}-E^{-}_{0})\eta^{-1}$ \\
                 $\eta^{-1}(E^{+}_{0}-E^{-}_{0})$ & $2\eta^{-1}$ \\
                 \end{tabular} \right).
\end{equation}
Note that this matrix is symmetric, i.e. assuming
$E'_{AB}=G'_{AB}+B'_{AB}$; then for this example $B'_{AB}=0$ and
consequently $H'_{ABC}=0$. Now by applying the formalism of the
previous section for the above example, we see that this is an
example of the cases d) or f); i.e. we must impose condition (41)-(43)
 for having N=(2,2) supersymmetry  on the action S (45).
\par The next step for investigating the invariance of N=(2,2)
supersymmetry structure under Poisson-Lie T-duality is of obtaining
the sigma model action (50) and its dual (51) by using the
decompositions $L=g\tilde{h}$ and $L=\tilde{g}h$ in the action (45)(with the
above restrictions on $G'$); then one can investigate the N=(2,2) structure
on these actions. Note that for these sigma models the background
matrix $E_{AB}(g)(\tilde{E}(\tilde{g}))$ is dependent on the Lie group coordinates
and one can not use the above algebraic formulation; and one must
control the conditions (4)-(6) and (9) directly for these
models.\\
\par \textbf{Acknowledgment:}
We would like to thanks from F.Darabi for useful comments.


\end{document}